\documentclass{article}
\usepackage{amssymb}
\usepackage{amsfonts}
\usepackage{amsmath}

\input{tcilatex}
\begin{document}

\author{J. G. Cardoso\thanks{%
jorge.cardoso@udesc.br} \\
Department of Mathematics\\
Centre for Technological Sciences-UDESC\\
Joinville 89223-100 SC, Brazil.
torsional affinities; dark energy, wave functions, wave equations.}
\title{A Local Description of Dark Energy in Terms of Classical
Two-Component Massive Spin-One Uncharged Fields on Spacetimes with
Torsionful Affinities}
\date{ }
\maketitle

\begin{abstract}
It is assumed that the two-component spinor formalisms for curved spacetimes
that are endowed with torsionful affine connexions can supply a local
description of dark energy in terms of classical massive spin-one uncharged
fields. The relevant wave functions are related to torsional affine
potentials which bear invariance under the action of the generalized Weyl
gauge group. Such potentials are thus taken to carry an observable character
and emerge from contracted spin affinities whose patterns are chosen in a
suitable way. New covariant calculational techniques are then developed
towards deriving explicitly the wave equations that supposedly control the
propagation in spacetime of the dark energy background. What immediately
comes out of this derivation is a presumably natural display of interactions
between the fields and both spin torsion and curvatures. The physical
properties that may arise directly from the solutions to the wave equations
are not brought out.
\end{abstract}

\section{Introduction}

Since the discovery of the cosmic dark energy [1, 2], several attempts have
been made [3-9] at accomplishing a macroscopic explanation of the presently
observable acceleration of the universe [10, 11], while circumventing the
situations concerning some of the problems that arise in the context of the
standard cosmology [4, 12]. One of the most popular approaches that were
designed in this connection describes dark energy in a geometrically
torsionless fashion as a gravitationally repulsive cosmic background
modelled either by a positive cosmological constant or by a scalar field to
which a physical meaning may possibly be ascribed. In this model, the dark
energy density can be explicitly evaluated with the help of some auxiliary
observational data, but the corresponding results nevertheless turn out to
be in serious disagreement with characteristic values arising from the
conventional quantum field theories. In addition, the complete physical
adequacy of the scalar field taken up thereabout has not been established
hitherto. Another popular approach focusses upon trivial modifications of
the Lagrangian density for classical general relativity. It likewise
implements alternative patterns for generally relativistic energy momentum
tensors, and thereby gives rise to the need for sorting out the microscopic
nature of dark energy within extended particle physics models. A somewhat
interesting work carried out along these lines [13], identifies the dark
energy background with a massive vector potential which is taken from the
beginning to obey a non-minimal coupling to gravity. Accordingly, the
Friedmann equations acquire an extra non-geometric term which is
proportional to the rest mass of the dark energy particles. Moreover, the
implementation of certain astronomical constraints makes it feasible to
estimate the mass of the particles. The overall picture then leads to a mass
value naively related to the cosmological constant, and also supplies a
late-time accelerated De Sitter-like cosmic expansion.

On the basis of Einstein-Cartan's theory [14-18], a prospect has been posed
by researchers for bringing forth a torsional version of the standard
cosmological model (see Refs. [19, 20]). This had been partially motivated
by a theoretical possibility of particularly explaining the cosmic
acceleration of the universe along with its spatial flatness, its
homogeneity and isotropy, without having to call for any mechanisms of
cosmic inflation [3, 4]. As mentioned in Refs. [21-25], torsional gravity
has also attracted a considerable deal of attention in conjunction with a
prediction achieved by string theory that concerns the occurrence of
couplings between torsion and spinning fields. Many insights into the
understanding of both the coupling strengths of the fundamental interactions
and the ratios between them, have thus been gained from the torsionic
property of underlying spacetime geometries. Remarkably enough, the
essentially unique torsionful version of the famous Infeld-van der Waerden $%
\gamma \varepsilon $-formalisms [26-38] had been until very recently [39]
just sparsely considered in the literature [40, 41]. The main motivation for
formulating this torsional extension came from the ascertainment that its
geometric inner structure may allow the implementation of affine
contributions which afford gauge invariant vector potentials bearing an
observable character. It had then been expected that the definitive
ascription of a fundamental significance to spacetime torsion would
eventually become more tangible if a torsional two-component spinor
description of dark energy might go hand-in-hand with the spin-torsion
mechanisms that prevent the universe from being originated by a singularity
[42-44].

In the present work, we take account of the torsional spinor formalisms
referred to previously to bring forward a supposedly realistic description
of the dynamics of dark energy in a purely local fashion. In fact, the
viability for carrying out our description relies geometrically upon the
possibility of choosing asymmetric spin-affine connexions that supply gauge
invariant potentials for two-component massive spin-one uncharged fields on
spacetimes with torsionful affinities. The paper works out the idea that the
universe could have been expected beforehand to host two physical
backgrounds which, as we believe, must be described in terms of affine
potentials coming from the spinor structures inherently borne by generally
relativistic spacetimes [45, 46]. Hence, a torsionless electromagnetic
background should be locally described by the old $\gamma \varepsilon $%
-formalisms such as suggested in Refs. [29, 33], and a torsionful background
should be describable locally in terms of geometric Proca fields within a
suitably extended spinor framework. Throughout the paper, we thus adopt the
attitude that identifies the former with the cosmic microwave background
(CMB), and likewise think of the latter as constituting the cosmic dark
energy. As was pointed out in Ref. [39] from a strictly geometric viewpoint,
any torsional affine potential must be accompanied by proper torsionless
contributions whence, in actuality, the implementation of this picture gives
rise to one of the theoretical features of our work whereby the spacetime
description of dark energy has to be united together with that of the CMB.
Yet, we realize that the propagation of the CMB in regions of the universe
where the values of torsional affinities are negligible may be described
alone within the framework of Ref. [28].

We shall account for the well-established observational fact [8, 9] that the
CMB and dark energy permeate together the whole of the universe. Because of
the locality of our description, the completion of the relevant procedures
will be accomplished without making it necessary to allow for any
cosmological kinematics or even to call upon any ordinary cosmological
presuppositions like those concerning homogeneities, isotropy, inflation and
shape of physical densities. Instead, the only assumptions lying behind the
implementation of our procedures are the same as the ones made before [39],
according to which local spinor structures along with manifold mapping
groups and the matrices that classically constitute the generalized Weyl
gauge group [26-28], remain all formally unaltered when any classical
spacetime consideration is shifted to the torsional framework. We stress
that the defining prescriptions for any of the geometric world and spin
densities tied in with the old formalisms [28, 29], may be applicable
equally well herein. The information on the wave functions for both physical
backgrounds is carried by adequately contracted spin curvatures which emerge
as sums of typical bivector contributions from the action on arbitrary spin
vectors of a characteristic torsionful second-order covariant derivative
operator. It appears that the additivity property of such contracted
curvatures is really passed on to the wave functions.

We will utilize the notation adhered to in Ref. [39]. Unless otherwise
indicated in an explicit manner, the usual designation of the traditional
spinor framework as $\gamma \varepsilon $-formalisms will henceforward be
attributed to the torsionful two-component formalisms under consideration
here. Upon writing down the world form of the pertinent field equations, we
shall therefore take into account geometric electromagnetic and uncharged
Proca fields for a curved spacetime $\mathfrak{M}$ that carries a world
metric tensor $g_{\mu \nu }$ having the local signature $(+---)$ and a
torsionful, metric compatible, covariant derivative operator $\nabla _{\mu }$%
. The spinor form of the field equations will be obtained by carrying out a
straightforward transcription of the respective world statements. We will
see that the resulting spinor field equations involve pairs of new complex
conjugate current densities for each physical background, which absorb outer
products carrying appropriate torsion spinors along with the wave functions
themselves. In order to carry out systematically the derivation of the wave
equations that control the propagation of the fields in $\mathfrak{M}$, we
shall have to adapt to the torsional framework the differential
calculational techniques employed for the first time in the work of Ref.
[28]. What immediately comes out of this derivation is a presumably natural
display of interactions between the fields and both torsion and curvatures.
In either formalism, some pieces of the geometric sources originated by the
field equations must thus be subject to prescribed gauge invariant
subsidiary conditions which are brought about by the inherent symmetry of
the wave functions. We will not bring out at this stage any physical
properties that may arise from the solutions to our wave equations, however.

Without any risk of confusion, we will use the same indexed symbol $\nabla
_{\mu }$ to write covariant derivatives in both formalisms. The symbol $%
\mathfrak{g}$ will sometimes be used for denoting the determinant of $g_{\mu
\nu }$. For the world affine connexion associated with $\nabla _{\mu }$, we
have the splitting%
\begin{equation*}
\Gamma _{\mu \nu \lambda }=\widetilde{\Gamma }_{\mu \nu \lambda }+T_{\mu \nu
\lambda },
\end{equation*}%
where $\widetilde{\Gamma }_{\mu \nu \lambda }=\Gamma _{(\mu \nu )\lambda }$
and $T_{\mu \nu \lambda }=\Gamma _{\lbrack \mu \nu ]\lambda }$ is by
definition the torsion tensor of $\nabla _{\mu }$. The symmetric piece $%
\widetilde{\Gamma }_{\mu \nu \lambda }$ may be identified with the
Christoffel connexion of $g_{\mu \nu }$ in case $T_{\mu \nu \lambda }$ is
rearranged adequately. We take the elements of the Weyl gauge group as
non-singular complex $(2\times 2)$-matrices whose entries are defined by%
\begin{equation*}
\Lambda _{A}{}^{B}=\exp (i\theta )\delta _{A}{}^{B},
\end{equation*}%
where $\delta _{A}{}^{B}$ denotes the Kronecker symbol and $\theta $ is the
gauge parameter of the group which shows up as an arbitrary differentiable
real-valued function on $\mathfrak{M}$. The determinant $\exp (2i\theta )$
of $(\Lambda _{A}{}^{B})$ will be denoted as $\Delta _{{\small \Lambda }}$.
A horizontal bar lying over some kernel letter will denote the operation of
complex conjugation. Some minor conventions shall be explained in due course.

Our outline has been set as follows. In Section 2, we recall the contracted
spin curvatures as built up in Ref. [39], and bring out the world field
equations. The definition of all wave functions is shown in Section 3
together with the spinor field equations. In Section 4, the torsional
calculational techniques are developed. There, we will have to consider spin
curvatures somewhat further. Nonetheless, many of the curvature formulae
deduced in Ref. [39] shall be taken for granted at the outset. In Section 5,
the wave equations are derived. We set an outlook on future works in Section
6.

\section{World Field Equations}

The key curvature object for either formalism is a world-spin quantity $%
C_{\mu \nu AB}{}$ that occurs in the configuration%
\begin{equation}
D_{\mu \nu }\zeta ^{B}=C_{\mu \nu A}{}^{B}\zeta ^{A},  \label{1}
\end{equation}%
where $\zeta ^{A}$ is an arbitrary spin vector and $D_{\mu \nu }$ amounts to
the characteristic second-order covariant derivative operator of the
torsional framework, namely,%
\begin{equation}
D_{\mu \nu }\doteqdot 2(\nabla _{\lbrack \mu }\nabla _{\nu ]}+T_{\mu \nu
}{}^{\lambda }\nabla _{\lambda }).  \label{2}
\end{equation}%
In the $\gamma $-formalism, we have the tensor law 
\begin{equation}
C_{\mu \nu AB}^{\prime }={}\Lambda _{A}{}^{C}\Lambda _{B}{}^{D}C_{\mu \nu
CD}=\Delta _{{\small \Lambda }}C_{\mu \nu AB},  \label{2Lin}
\end{equation}%
whereas the object $C_{\mu \nu AB}{}$ for the $\varepsilon $-formalism is
taken as an invariant spin-tensor density of weight $-1$, that is to say,%
\begin{equation}
C_{\mu \nu AB}^{\prime }{}=(\Delta _{{\small \Lambda }})^{-1}\Lambda
_{A}{}^{C}\Lambda _{B}{}^{D}C_{\mu \nu CD}=C_{\mu \nu AB}.  \label{3Lin}
\end{equation}

The contracted curvature $C_{\mu \nu A}{}^{A}{}$ possesses the gauge
invariant additivity property\footnote{%
We should emphasize that the uncontracted object $C_{\mu \nu A}{}^{B}$ for
either formalism does \textit{not }hold the additivity property.}%
\begin{equation}
C_{\mu \nu A}{}^{A}=\widetilde{C}_{\mu \nu A}{}^{A}+C_{\mu \nu A}^{%
{\footnotesize (T)}}{}^{A}.  \label{3}
\end{equation}%
In particular, $C_{\mu \nu A}^{{\footnotesize (T)}}{}^{A}$ accounts for the
torsionfulness of $\nabla _{\mu }$ while the whole $\widetilde{C}_{\mu \nu
AB}{}$ is taken up by the torsionless commutator%
\begin{equation}
2\widetilde{\nabla }_{[\mu }\widetilde{\nabla }_{\nu ]}\zeta ^{B}=\widetilde{%
C}_{\mu \nu A}{}^{B}\zeta ^{A},  \label{4}
\end{equation}%
where $\widetilde{\nabla }_{\mu }$ is indeed the covariant derivative
operator for $\widetilde{\Gamma }_{\mu \nu \lambda }$. It turns out that we
can write down the simultaneous contracted relations%
\begin{equation}
\widetilde{C}_{\mu \nu A}{}^{A}=2\partial _{\lbrack \mu }\widetilde{%
\vartheta }_{\nu ]A}{}^{A},\text{ }C_{\mu \nu A}^{{\footnotesize (T)}%
}{}^{A}=2\partial _{\lbrack \mu }\vartheta _{\nu ]A}^{{\footnotesize (T)}%
}{}^{A},  \label{5}
\end{equation}%
with the involved $\vartheta $-pieces thus occurring in the skew
contributions that make up in each formalism a suitably chosen asymmetric
spin affinity for $\nabla _{\mu }$, in agreement with Eq. (\ref{3}). Hence,
making use of the standard patterns [39]%
\begin{equation}
\widetilde{\vartheta }_{\mu A}{}^{A}=\partial _{\mu }\log E-2i\Phi _{\mu },%
\text{ }\vartheta _{\mu A}^{{\footnotesize (T)}}{}^{A}=-2iA_{\mu },
\label{6}
\end{equation}%
yields the purely imaginary expression%
\begin{equation}
C_{\mu \nu A}{}^{A}=-2i(\widetilde{F}_{\mu \nu }+F_{\mu \nu }^{%
{\footnotesize (T)}}),  \label{7}
\end{equation}%
along with the bivectors%
\begin{equation}
\widetilde{F}_{\mu \nu }\doteqdot 2\partial _{\lbrack \mu }\Phi _{\nu ]},%
\text{ }F_{\mu \nu }^{{\footnotesize (T)}}\doteqdot 2\partial _{\lbrack \mu
}A_{\nu ]},  \label{8}
\end{equation}%
with $\Phi _{\mu }$ and $A_{\mu }$ amounting to affine potentials subject to
the gauge behaviours%
\begin{equation}
\Phi _{\mu }^{\prime }=\Phi _{\mu }-\partial _{\mu }\theta ,\text{ }A_{\mu
}^{\prime }=A_{\mu }.  \label{9}
\end{equation}%
It is worthwhile to recast each of the derivatives of Eq. (\ref{8}) as a
piece that looks formally like%
\begin{equation}
\partial _{\lbrack \mu }\Omega _{\nu ]}=\nabla _{\lbrack \mu }\Omega _{\nu
]}+T_{\mu \nu }{}^{\lambda }\Omega _{\lambda }.  \label{10}
\end{equation}

We mention, in passing, that the quantity $E$ carried by the prescriptions (%
\ref{6}) is a real positive-definite world-invariant spin-scalar density of
absolute weight $+1$. In the $\gamma $-formalism, it carries a manifestly
spin-metric character, but this ceases holding for the $\varepsilon $%
-formalism. The potentials $\Phi _{\mu }$ and $A_{\mu }$ are the same in
both formalisms. They arise from an affine property of the covariant
derivative expansions for the Hermitian connecting objects of the formalisms
(for further details, see Ref. [39]).

It can be seen from Eq. (\ref{9}) that $\Phi _{\mu }$ is a Maxwell
potential, which we take to be physically associated to the CMB. In turn, $%
A_{\mu }$ bears gauge invariance and is likewise looked upon as a potential
of mass $m$ for the dark energy background. The world form of the first half
of the overall set of field equations emerges from the usual least-action
principles for Maxwell and real Proca fields in curved spacetimes [47]. It
follows that, allowing for the relation%
\begin{equation}
\frac{1}{\sqrt{-\mathfrak{g}}}\partial _{\mu }(\sqrt{-\mathfrak{g}}%
\tciFourier ^{\mu \lambda })=\nabla _{\mu }\tciFourier ^{\mu \lambda
}+2T_{\mu }\tciFourier ^{\mu \lambda }-T_{\mu \nu }{}^{\lambda }\tciFourier
^{\mu \nu },  \label{11}
\end{equation}%
with $T_{\mu }\doteqdot T_{\mu \tau }{}^{\tau }$ and the kernel letter $%
\tciFourier $ standing for either $\widetilde{F}$ or $F^{{\footnotesize (T)}%
} $, we get the first half of Maxwell's equations 
\begin{equation}
\nabla ^{\mu }\widetilde{F}_{\mu \lambda }+2T^{\mu }\widetilde{F}_{\mu
\lambda }-T{}^{\mu \nu }{}_{\lambda }\widetilde{F}_{\mu \nu }=0,\text{ }
\label{12}
\end{equation}%
along with the first half of Proca's equations%
\begin{equation}
\nabla ^{\mu }F_{\mu \lambda }^{{\footnotesize (T)}}+2T^{\mu }F_{\mu \lambda
}^{{\footnotesize (T)}}-T{}^{\mu \nu }{}_{\lambda }F_{\mu \nu }^{%
{\footnotesize (T)}}+m^{2}A_{\lambda }=0.  \label{13}
\end{equation}%
Obviously, in accordance with our picture, the statements (\ref{12}) and (%
\ref{13}) are the dynamical world field equations in $\mathfrak{M}$ for CMB
photons and dark energy fields. Both of the second halves come about as the
corresponding Bianchi identities, which may be expressed by%
\begin{equation}
\nabla ^{\mu }{}^{\ast }\tciFourier _{\mu \lambda }=-2{}^{\ast }T_{\lambda
}{}^{\mu \nu }\tciFourier _{\mu \nu },  \label{14}
\end{equation}%
with the kernel-letter notation of (\ref{11}), as well as some of the
dualization schemes given in Ref. [16], having been utilized for writing Eq.
(\ref{14}).

\section{Spinor Field Equations}

The wave functions for both backgrounds are supplied by the spinor
decomposition of the bivectors carried by Eq. (\ref{8}). We have, in effect,%
\begin{equation}
S_{AA^{\prime }}^{\mu }S_{BB^{\prime }}^{\nu }\widetilde{F}_{\mu \nu
}=M_{A^{\prime }B^{\prime }}\phi _{AB}+M_{AB}\phi _{A^{\prime }B^{\prime }}
\label{15}
\end{equation}%
and\footnote{%
The kernel letter $M$ will henceforth denote either $\gamma $ or $%
\varepsilon $.}%
\begin{equation}
S_{AA^{\prime }}^{\mu }S_{BB^{\prime }}^{\nu }F_{\mu \nu }^{{\footnotesize %
(T)}}=M_{A^{\prime }B^{\prime }}\psi _{AB}+M_{AB}\psi _{A^{\prime }B^{\prime
}},  \label{16}
\end{equation}%
where the $S$-symbols are some of the connecting objects for the formalism
occasionally allowed for, and the entries of the pair $(M_{AB},M_{A^{\prime
}B^{\prime }})$ just denote the respective covariant metric spinors. Thus,
the wave functions carried by $(\phi _{AB},\phi _{A^{\prime }B^{\prime }})$
and $(\psi _{AB},\psi _{A^{\prime }B^{\prime }})$ come into play as massless
and massive spin-one uncharged fields of opposite handednesses, with their
gauge characterizations incidentally coinciding with those exhibited by Eqs.
(\ref{2Lin}) and (\ref{3Lin}). By invoking Eq. (\ref{10}) together with the
torsion decomposition%
\begin{equation}
T_{AA^{\prime }BB^{\prime }}{}^{\mu }=M_{A^{\prime }B^{\prime }}\tau
_{AB}{}^{\mu }+M_{AB}\tau _{A^{\prime }B^{\prime }}{}^{\mu },  \label{17}
\end{equation}%
we obtain the field-potential relationships%
\begin{equation}
\phi _{AB}{}=-\nabla _{(A}^{C^{\prime }}\Phi _{B)C^{\prime }}+2\tau
_{AB}{}^{\mu }\Phi _{\mu }  \label{18}
\end{equation}%
and%
\begin{equation}
\psi _{AB}=-\nabla _{(A}^{C^{\prime }}A_{B)C^{\prime }}+2\tau _{AB}{}^{\mu
}A_{\mu }.  \label{19}
\end{equation}%
The contravariant form of (\ref{18}) and (\ref{19}) is written in both
formalisms as%
\begin{equation}
\phi ^{AB}=\nabla _{C^{\prime }}^{(A}\Phi ^{B)C^{\prime }}+2\tau ^{AB\mu
}\Phi _{\mu }  \label{18Lin}
\end{equation}%
and%
\begin{equation}
\psi ^{AB}=\nabla _{C^{\prime }}^{(A}A^{B)C^{\prime }}+2\tau ^{AB\mu }A_{\mu
},  \label{19Lin}
\end{equation}%
where we have implemented the eigenvalue equations%
\begin{equation}
\nabla _{\mu }\gamma _{AB}=i\alpha _{\mu }\gamma _{AB},\text{ }\nabla _{\mu
}\gamma ^{AB}=-i\alpha _{\mu }\gamma ^{AB},  \label{22}
\end{equation}%
together with their conjugates and the definition%
\begin{equation}
\alpha _{\mu }\doteqdot \partial _{\mu }\Phi +2(\Phi _{\mu }+A_{\mu }),
\label{Ad22}
\end{equation}%
with the quantity $\Phi $ being nothing else but the polar argument of the
independent component of $\gamma _{AB}$ (see Eq. (\ref{add1}) below).

We next carry out the spinor translation of the individual pieces of Eqs. (%
\ref{12})-(\ref{14}), with the purpose of paving the way for deriving the
field equations at issue. Evidently, it will suffice to carry through the
apposite procedures for either of the $\tciFourier $-bivectors of Eq. (\ref%
{11}). For the $\nabla $-term of (\ref{13}), say, we have%
\begin{equation}
\nabla ^{AA^{\prime }}F_{AA^{\prime }BB^{\prime }}^{{\footnotesize (T)}%
}=\nabla ^{AA^{\prime }}(M_{A^{\prime }B^{\prime }}\psi _{AB})+\text{c.c.},
\label{20}
\end{equation}%
with the symbol "c.c." denoting an overall complex conjugate piece. In the $%
\gamma $-formalism, the right-hand side of Eq. (\ref{20}) reads%
\begin{equation}
\nabla ^{AA^{\prime }}(\gamma _{A^{\prime }B^{\prime }}\psi _{AB})+\text{c.c.%
}=(\nabla _{B^{\prime }}^{A}\psi _{AB}-i\alpha _{B^{\prime }}^{A}\psi _{AB})+%
\text{c.c.}.  \label{21}
\end{equation}%
As $\nabla _{\mu }\varepsilon _{AB}=0$ in both formalisms, the $\varepsilon $%
-formalism counterpart of (\ref{21}) may be obtained by dropping the $\alpha 
$-term from it. By combining (\ref{16}) and (\ref{17}), we readily find the
patterns%
\begin{equation}
T^{AA^{\prime }}F_{AA^{\prime }BB^{\prime }}^{{\footnotesize (T)}}=(\tau
^{AM}{}_{MB^{\prime }}-\tau _{B^{\prime }M^{\prime }}{}^{AM^{\prime }})\psi
_{AB}+\text{c.c.}  \label{23}
\end{equation}%
and%
\begin{equation}
T^{AA^{\prime }MM^{\prime }}{}_{BB^{\prime }}F_{AA^{\prime }MM^{\prime }}^{%
{\footnotesize (T)}}=2\tau ^{AM}{}_{BB^{\prime }}\psi _{AM}+\text{c.c.},
\label{24}
\end{equation}%
which just represent $T^{\mu }F_{\mu \lambda }^{{\footnotesize (T)}}$ and $%
T{}^{\mu \nu }{}_{\lambda }F_{\mu \nu }^{{\footnotesize (T)}}$ in either
formalism. The $\gamma $-formalism version of the left-hand side of Eq. (\ref%
{14}) is given by%
\begin{equation}
\nabla ^{AA^{\prime }}{}^{\ast }F_{AA^{\prime }BB^{\prime }}^{{\footnotesize %
(T)}}=i[(\nabla _{B}^{A^{\prime }}\psi _{A^{\prime }B^{\prime }}+i\alpha
_{B}^{A^{\prime }}\psi _{A^{\prime }B^{\prime }})-\text{c.c.}],  \label{25}
\end{equation}%
whereas the piece $^{\ast }T_{\lambda }{}^{\mu \nu }F_{\mu \nu }^{%
{\footnotesize (T)}}$ gets in each formalism translated into%
\begin{equation}
^{\ast }T_{BB^{\prime }}{}^{AA^{\prime }MM^{\prime }}F_{AA^{\prime
}MM^{\prime }}^{{\footnotesize (T)}}=i[(\tau _{B}{}^{AM}{}_{B^{\prime }}\psi
_{AM}-\text{c.c.})+(\tau _{B^{\prime }M^{\prime }}{}^{AM^{\prime }}\psi
_{AB}-\text{c.c.})].  \label{26}
\end{equation}

Towards completing our derivation procedures, it is convenient to require
the unprimed and primed wave functions for either background to bear
algebraic independence throughout $\mathfrak{M}$. This requirement enables
us to manipulate the configurations involved in the spinor transcription we
have carried out above in such a way that the left-right handedness
characters of the fields become transparently separable. Therefore, by
taking into account the equality%
\begin{equation}
\tau ^{AM}{}_{MB^{\prime }}-\tau _{B^{\prime }M^{\prime }}{}^{AM^{\prime
}}=T_{B^{\prime }}^{A},  \label{27}
\end{equation}%
we obtain the field equation%
\begin{equation}
\nabla ^{AA^{\prime }}(M_{A^{\prime }B^{\prime }}\psi _{AB})+\frac{1}{2}%
m^{2}A_{BB^{\prime }}=s_{BB^{\prime }},  \label{29}
\end{equation}%
with the complex dark energy source%
\begin{equation}
s_{BB^{\prime }}=2(\tau ^{AM}{}_{BB^{\prime }}\psi _{AM}-T_{B^{\prime
}}^{A}\psi _{AB}).  \label{29Lin}
\end{equation}%
It should be remarked that the term $\tau _{B}{}^{AM}{}_{B^{\prime }}\psi
_{AM}$, which is borne by Eq. (\ref{26}), cancels out at an intermediate
step of the manipulations that yield the statement (\ref{29}), and thence
also so does its complex conjugate. In the $\gamma $-formalism, we then have%
\begin{equation}
\nabla _{B^{\prime }}^{A}\psi _{AB}-i\alpha _{B^{\prime }}^{A}\psi _{AB}+%
\frac{1}{2}m^{2}A_{BB^{\prime }}=s_{BB^{\prime }},  \label{31}
\end{equation}%
with the corresponding $\varepsilon $-formalism statement being spelt out as%
\begin{equation}
\nabla _{B^{\prime }}^{A}\psi _{AB}+\frac{1}{2}m^{2}A_{BB^{\prime
}}=s_{BB^{\prime }}.  \label{32Lin}
\end{equation}%
For the CMB, we get the $\gamma $-formalism massless field equation%
\begin{equation}
\nabla _{B^{\prime }}^{A}\phi _{AB}-i\alpha _{B^{\prime }}^{A}\phi _{AB}=%
\mathfrak{s}_{BB^{\prime }},  \label{30}
\end{equation}%
along with its $\varepsilon $-formalism counterpart%
\begin{equation}
\nabla _{B^{\prime }}^{A}\phi _{AB}=\mathfrak{s}_{BB^{\prime }}  \label{32}
\end{equation}%
and the geometric source%
\begin{equation}
\mathfrak{s}_{BB^{\prime }}=2(\tau ^{AM}{}_{BB^{\prime }}\phi
_{AM}-T_{B^{\prime }}^{A}\phi _{AB}).  \label{32Linn}
\end{equation}

It was demonstrated in Ref. [28] that the wave-function pattern $\phi
_{A}{}^{B}$ for the torsionless framework bears a commonness feature in that
it is the same in both the classical formalisms. Inasmuch as the traditional
algebraic definitions for metric spinors and connecting objects are formally
appropriate for the torsionful framework as well, we can right away write
the $\gamma \varepsilon $-relationships%
\begin{equation}
C_{\mu \nu A}^{(\gamma )}{}^{B}=C_{\mu \nu A}^{(\varepsilon
)}{}^{B}\Leftrightarrow C_{\mu \nu AB}^{(\gamma )}{}=\gamma C_{\mu \nu
AB}^{(\varepsilon )}{},  \label{add1}
\end{equation}%
where $\gamma $ is the independent component of $\gamma _{AB}$. Consequently,%
\footnote{%
We will henceforth stop staggering the indices of any symmetric two-index
configuration.} we can say that each of the pairs $(\phi _{A}^{B}{}{},\phi
_{A^{\prime }}^{B^{\prime }}{}{})$ and $(\psi _{A}^{B}{}{},\psi _{A^{\prime
}}^{B^{\prime }}{}{})$ possesses a commonness property which is seemingly
similar to the classical one, in addition to holding in both formalisms a
gauge invariant spin-tensor character. In each formalism, we thus have the
field equations%
\begin{equation}
\nabla ^{AB^{\prime }}\psi _{A}^{B}{}+\frac{1}{2}m^{2}A^{BB^{\prime
}}=s^{BB^{\prime }}  \label{add3}
\end{equation}%
and%
\begin{equation}
\nabla ^{AB^{\prime }}\phi _{A}^{B}{}=\mathfrak{s}^{BB^{\prime }},
\label{add2}
\end{equation}%
where the $\phi $-field relation, for instance,%
\begin{equation}
\gamma _{CB}\nabla ^{AB^{\prime }}\phi _{A}^{C}{}=\nabla ^{AB^{\prime }}\phi
_{AB}-i\alpha ^{AB^{\prime }}\phi _{AB},  \label{add9}
\end{equation}%
has been used in the $\gamma $-formalism case.

\section{Calculational Techniques}

By this point, we shall build up the techniques that yield in both
formalisms the wave equations for the fields being considered. In fact,
these techniques constitute a torsional version of the differential ones
which had been developed originally within the classical $\gamma \varepsilon 
$-framework [28]. Let us begin with the operator decomposition%
\begin{equation}
S_{AA^{\prime }}^{\mu }S_{BB^{\prime }}^{\nu }D_{\mu \nu }=M_{A^{\prime
}B^{\prime }}\check{D}_{AB}+M_{AB}\check{D}_{A^{\prime }B^{\prime }}.
\label{34}
\end{equation}%
Whence, implementing Eqs. (\ref{2}) and (\ref{17}), gives%
\begin{equation}
\check{D}_{AB}=\Delta _{AB}+2\tau _{AB}{}^{\mu }\nabla _{\mu },\text{ }%
\Delta _{AB}\doteqdot -\nabla _{(A}^{C^{\prime }}\nabla _{B)C^{\prime }},
\label{35}
\end{equation}%
together with the complex conjugate of (\ref{35}). The operators $\check{D}%
_{AB}$ and $\Delta _{AB}$\ both are linear and possess the Leibniz rule
property.

It may be useful to utilize Eq. (\ref{22}) for reexpressing the $\gamma $%
-formalism operator $\Delta _{AB}$ as%
\begin{equation}
\Delta _{AB}=\nabla _{C^{\prime }(A}\nabla _{B)}^{C^{\prime }}-i\alpha
_{C^{\prime }(A}\nabla _{B)}^{C^{\prime }}.  \label{36}
\end{equation}%
In the $\varepsilon $-formalism, one has%
\begin{equation}
\Delta _{AB}=-\nabla _{(A}^{C^{\prime }}\nabla _{B)C^{\prime }}=\nabla
_{C^{\prime }(A}\nabla _{B)}^{C^{\prime }}.  \label{37}
\end{equation}%
It is worth noticing that the $\gamma $-formalism contravariant form of $%
\Delta _{AB}$ appears as%
\begin{equation}
\Delta ^{AB}=-(\nabla ^{C^{\prime }(A}\nabla _{C^{\prime }}^{B)}+i\alpha
^{C^{\prime }(A}\nabla _{C^{\prime }}^{B)}),  \label{38}
\end{equation}%
or, equivalently, as 
\begin{equation}
\Delta ^{AB}=\nabla _{C^{\prime }}^{(A}\nabla ^{B)C^{\prime }}.  \label{39}
\end{equation}%
Because $\alpha _{\mu }$ bears gauge invariance [39], the conjugate $\check{D%
}$-operators for the $\gamma $-formalism behave under gauge transformations
as covariant spin tensors. In the $\varepsilon $-formalism, they
correspondingly behave as invariant spin-tensor densities of weight $-1$ and
antiweight $-1$.

Equations (\ref{1}) and (\ref{34}) suggest that some of the elementary $%
\check{D}$-derivatives should be prescribed in either formalism by 
\begin{equation}
\check{D}_{AB}\zeta ^{C}=\varpi _{ABM}{}^{C}\zeta ^{M},\text{ }\check{D}%
_{A^{\prime }B^{\prime }}\zeta ^{C}=\varpi _{A^{\prime }B^{\prime
}M}{}^{C}\zeta ^{M},  \label{40}
\end{equation}%
with the spin-curvature expansion%
\begin{equation}
C_{AA^{\prime }BB^{\prime }CD}=M_{A^{\prime }B^{\prime }}\varpi
_{ABCD}+M_{AB}\varpi _{A^{\prime }B^{\prime }CD},  \label{40Lin}
\end{equation}%
and the relationships%
\begin{equation}
\varpi _{ABCD}^{(\gamma )}=\gamma ^{2}\varpi _{ABCD}^{(\varepsilon )},\text{ 
}\varpi _{A^{\prime }B^{\prime }CD}^{(\gamma )}=\mid \gamma \mid ^{2}\varpi
_{A^{\prime }B^{\prime }CD}^{(\varepsilon )},  \label{add4}
\end{equation}%
which clearly agree with (\ref{add1}). We can show [39] that the spinor pair%
\begin{equation}
\mathbf{G}=(\varpi _{AB(CD)}{},\varpi _{A^{\prime }B^{\prime }(CD)}{})
\label{addG}
\end{equation}%
constitutes the irreducible decomposition of the Riemann tensor for $\nabla
_{\mu }$. Its unprimed entry is expandable as\footnote{%
From now on, we will for simplicity employ the definitions X$%
_{ABCD}\doteqdot \varpi _{AB(CD)}{}$ and $\Xi _{A^{\prime }B^{\prime
}CD}{}\doteqdot \varpi _{A^{\prime }B^{\prime }(CD)}$.}%
\begin{equation}
\text{X}_{ABCD}\hspace{-0.07cm}=\hspace{-0.07cm}\Psi _{ABCD}-M_{(A\mid
(C}\xi _{D)\mid B)}-\frac{1}{3}\varkappa M_{A(C}M_{D)B},  \label{41}
\end{equation}%
with%
\begin{equation}
\hspace{-0.07cm}\Psi _{ABCD}=\text{X}_{(ABCD)}\hspace{-0.07cm},\text{ }\xi
_{AB}=\text{X}^{M}{}_{(AB)M},\text{ }\varkappa =\text{X}_{LM}{}^{LM},
\label{42}
\end{equation}%
and the $\Psi $-spinor defining a typical wave function for gravitons in $%
\mathfrak{M}$. Likewise, the contracted pieces $(\varpi _{ABM}{}^{M},\varpi
_{A^{\prime }B^{\prime }M}{}^{M})$ fulfill the additivity relations (\ref{3}%
) and (\ref{7}), and are thereby proportional to the wave functions of (\ref%
{15}) and (\ref{16}) according to the schemes%
\begin{equation}
\widetilde{\varpi }_{ABM}{}^{M}=-2i\phi _{AB},\text{ }\widetilde{\varpi }%
_{A^{\prime }B^{\prime }M}{}^{M}=-2i\phi _{A^{\prime }B^{\prime }}
\label{43}
\end{equation}%
and%
\begin{equation}
\varpi _{ABM}^{{\footnotesize (T)}}{}^{M}=-2i\psi _{AB},\text{ }\varpi
_{A^{\prime }B^{\prime }M}^{{\footnotesize (T)}}{}^{M}=-2i\psi _{A^{\prime
}B^{\prime }}.  \label{44}
\end{equation}%
Hence, we can cast the prescriptions (\ref{40}) into the form%
\begin{equation}
\check{D}_{AB}\zeta ^{C}=\text{X}_{ABM}{}^{C}\zeta ^{M}-i(\phi _{AB}+\psi
_{AB})\zeta ^{C}  \label{45}
\end{equation}%
and 
\begin{equation}
\check{D}_{A^{\prime }B^{\prime }}\zeta ^{C}=\Xi _{A^{\prime }B^{\prime
}M}{}^{C}\zeta ^{M}-i(\phi _{A^{\prime }B^{\prime }}+\psi _{A^{\prime
}B^{\prime }})\zeta ^{C}.  \label{46}
\end{equation}

The prescriptions for computing $\check{D}$-derivatives of a covariant spin
vector $\eta _{A}$ can be obtained out of (\ref{40}) by assuming that%
\begin{equation}
\check{D}_{AB}(\zeta ^{C}\eta _{C})=0,\text{ }\check{D}_{A^{\prime
}B^{\prime }}(\zeta ^{C}\eta _{C})=0,  \label{47}
\end{equation}%
and carrying out Leibniz expansions thereof. We thus have 
\begin{equation}
\check{D}_{AB}\eta _{C}=-\hspace{1pt}[\text{X}_{ABC}{}^{M}\eta _{M}-i(\phi
_{AB}+\psi _{AB})\eta _{C}]  \label{48}
\end{equation}%
and%
\begin{equation}
\check{D}_{A^{\prime }B^{\prime }}\eta _{C}=-\hspace{1pt}[\Xi _{A^{\prime
}B^{\prime }C}{}^{M}{}\eta _{M}-i(\phi _{A^{\prime }B^{\prime }}+\psi
_{A^{\prime }B^{\prime }})\eta _{C}],  \label{49}
\end{equation}%
along with the complex conjugates of Eqs. (\ref{45})-(\ref{49}). The $\check{%
D}$-derivatives of a differentiable complex spin-scalar density $\alpha $ of
weight $\mathfrak{w}$ on $\mathfrak{M}$ are written out explicitly as%
\begin{equation}
\check{D}_{AB}\alpha =2i\mathfrak{w}\alpha (\phi _{AB}+\psi _{AB}),\text{ }%
\check{D}_{A^{\prime }B^{\prime }}\alpha =2i\mathfrak{w}\alpha (\phi
_{A^{\prime }B^{\prime }}+\psi _{A^{\prime }B^{\prime }}).  \label{50}
\end{equation}%
These configurations may in both formalisms be regarded as immediate
consequences of the integrability condition%
\begin{equation}
D_{\mu \nu }\alpha =2i\mathfrak{w}\alpha (\widetilde{F}_{\mu \nu }+F_{\mu
\nu }^{{\footnotesize (T)}}).  \label{51}
\end{equation}%
When acting on a world-spin scalar $h$, the $\check{D}$-operators then
recover the defining relation $D_{\mu \nu }h=0$ as%
\begin{equation}
\check{D}_{AB}h=0,\text{ }\check{D}_{A^{\prime }B^{\prime }}h=0,  \label{52}
\end{equation}%
whence%
\begin{equation}
\Delta _{AB}h=-2\tau _{AB}{}^{\mu }\nabla _{\mu }h.  \label{52Lin}
\end{equation}%
Of course, the patterns for $\check{D}$-derivatives of some spin-tensor
density can be specified from Leibniz expansions like 
\begin{equation}
\check{D}_{AB}(\alpha B_{C...D})=(\check{D}_{AB}\alpha )B_{C...D}+\alpha 
\check{D}_{AB}B_{C...D},  \label{53}
\end{equation}%
with $B_{C...D}$ being a spin tensor.

As for the old $\gamma \varepsilon $-framework, whenever $\check{D}$%
-derivatives of Hermitian quantities are actually computed in either
formalism, the wave function contributions carried by the expansions (\ref%
{45})-(\ref{49}) get cancelled. Such a cancellation likewise happens when we
let the $\check{D}$-operators act freely upon spin tensors having the same
numbers of covariant and contravariant indices of the same kind. For $%
\mathfrak{w}<0$, it still occurs in the expansion (\ref{53}) when $B_{C...D}$
is taken to carry $-2\mathfrak{w}$ indices and $\func{Im}\alpha \neq 0$
everywhere. A similar property also holds for situations that involve outer
products between contravariant spin tensors and complex spin-scalar
densities having suitable positive weights. The gauge behaviours specified
in the foregoing Section tell us that such weight-valence properties neatly
fit in with the case of the $\varepsilon $-formalism wave functions.

In carrying out the procedures for deriving our wave equations, it may
become necessary to take account of the algebraic rules%
\begin{equation}
2\nabla _{\lbrack B}^{A^{\prime }}\nabla _{A]A^{\prime }}=M_{AB}\square
=\nabla _{C}^{A^{\prime }}(M_{BA}\nabla _{A^{\prime }}^{C})  \label{54}
\end{equation}%
and 
\begin{equation}
2\nabla _{A^{\prime }}^{[A}\nabla ^{B]A^{\prime }}=M^{AB}\square =\nabla
_{A^{\prime }}^{C}(M^{BA}\nabla _{C}^{A^{\prime }}),  \label{55}
\end{equation}%
along with the operator splittings%
\begin{equation}
\nabla _{A}^{C^{\prime }}\nabla _{BC^{\prime }}=\frac{1}{2}M_{BA}\square
-\Delta _{AB},\text{ }\nabla _{A^{\prime }}^{A}\nabla ^{BA^{\prime }}=\Delta
^{AB}+\frac{1}{2}M^{AB}\square  \label{56}
\end{equation}%
and the gauge invariant definition 
\begin{equation}
\square \doteqdot \nabla _{AA^{\prime }}\nabla ^{AA^{\prime }}.  \label{57}
\end{equation}%
Owing to the applicability in both formalisms of the metric compatibility
condition%
\begin{equation}
\nabla _{\mu }(M_{AB}M_{A^{\prime }B^{\prime }})=0,  \label{58}
\end{equation}%
we can reset (\ref{57}) as%
\begin{equation}
\square =\nabla ^{AA^{\prime }}\nabla _{AA^{\prime }}.  \label{59}
\end{equation}%
In addition, from the equations%
\begin{equation}
\square \gamma _{AB}=\Theta \gamma _{AB},\text{ }\square \gamma ^{AB}=%
\overline{\Theta }\gamma ^{AB},  \label{60}
\end{equation}%
whose derivation involves using the eigenvalue carried by (\ref{22})
together with%
\begin{equation}
\Theta \doteqdot -\alpha ^{\mu }\alpha _{\mu }+i\nabla _{\mu }\alpha ^{\mu },
\label{61}
\end{equation}%
we also get the symbolic $\gamma $-formalism devices%
\begin{equation}
(\square \iota _{A}{}^{C})\gamma _{CB}=(\square -2i\alpha ^{\mu }\nabla
_{\mu }+\overline{\Theta })\iota _{AB}{}  \label{62}
\end{equation}%
and%
\begin{equation}
\gamma ^{AC}(\square \iota _{C}{}^{B})=(\square +2i\alpha ^{\mu }\nabla
_{\mu }+\Theta )\iota ^{AB},  \label{63}
\end{equation}%
which obey the valence-interchange rule\footnote{%
The rule (\ref{64}) had also arisen in Ref. [28] in connection with the
derivation of the wave equations for the CMB and gravitons in torsionless
environments.
\par
{}}%
\begin{equation}
i\alpha ^{\mu }\nabla _{\mu }\leftrightarrow -i\alpha ^{\mu }\nabla _{\mu },%
\text{ }\Theta \leftrightarrow \overline{\Theta }.  \label{64}
\end{equation}%
In the $\gamma $-formalism, the $\square $-correlations for $\iota _{AB}$
and $\iota ^{AB}$ can then be achieved from%
\begin{equation}
\gamma _{AC}\gamma _{BD}\square \iota ^{CD}=(\square -4i\alpha ^{\mu }\nabla
_{\mu }-\Upsilon )\iota _{AB}  \label{65}
\end{equation}%
and%
\begin{equation}
\gamma ^{AC}\gamma ^{BD}\square \iota _{CD}=(\square +4i\alpha ^{\mu }\nabla
_{\mu }-\overline{\Upsilon })\iota ^{AB},  \label{66}
\end{equation}%
which conform to Eq. (\ref{64}) with $\Upsilon =2(\alpha ^{\mu }\alpha _{\mu
}-\overline{\Theta })$.

\section{Wave Equations}

To obtain the entire set of wave equations that govern the propagation of
both physical backgrounds in $\mathfrak{M}$, we initially follow up the
simpler procedure which consists in implementing the calculational
techniques exhibited anteriorly to work out the field equation of either
formalism for the common dark energy wave function $\psi _{A}^{B}{}$. It
will become manifest that a gauge invariant condition for each entry of the
pairs $(\psi _{A}^{B}{},\psi _{A^{\prime }}^{B^{\prime }}{})$ and $(\phi
_{A}^{B}{},\phi _{A^{\prime }}^{B^{\prime }}{})$ can be established as a
geometric consequence of the symmetry of the underlying fields. Rather than
elaborating upon Eq. (\ref{31}), which could unnecessarily produce some
complicated manipulations, we will deduce the $\gamma $-formalism wave
equations for the unprimed pair $(\psi _{AB}{},\psi ^{AB}{})$ by appealing
to the differential devices (\ref{62}) and (\ref{63}). We may certainly get
the wave equations for any primed $\psi $-fields by taking complex
conjugates. The wave equations for all $\phi $-fields shall then arise in a
trivial way, provided that the field equations for both backgrounds carry
formally the same couplings between the wave functions and torsion spinors.

We start by operating with $\nabla _{B^{\prime }}^{C}$ on the configuration
of Eq. (\ref{add3}). Thus, recalling the contravariant splitting of (\ref{56}%
) leads us to the statement%
\begin{equation}
\Delta ^{AC}\psi _{A}^{B}{}-\frac{1}{2}M^{AC}\square \psi _{A}^{B}{}+\frac{1%
}{2}m^{2}\nabla _{B^{\prime }}^{C}A^{BB^{\prime }}=\nabla _{B^{\prime
}}^{C}s^{BB^{\prime }}.  \label{67}
\end{equation}%
It is obvious that both first-order derivative kernels of (\ref{67}) are of
the type%
\begin{equation}
\nabla _{B^{\prime }}^{C}u^{BB^{\prime }}=\nabla _{B^{\prime
}}^{(B}u^{C)B^{\prime }}-\frac{1}{2}M^{BC}\nabla _{\mu }u^{\mu },  \label{68}
\end{equation}%
with the symmetric piece for the potential being given by%
\begin{equation}
\nabla _{B^{\prime }}^{(B}A^{C)B^{\prime }}=\psi ^{BC}-2\tau ^{BC\mu }A_{\mu
},  \label{69}
\end{equation}%
in accordance with (\ref{19Lin}). By virtue of the relation (\ref{35}), the $%
\Delta $-piece of (\ref{67}) may be rewritten in either formalism as%
\begin{equation}
\Delta ^{AC}\psi _{A}^{B}{}=\check{D}^{AC}\psi _{A}^{B}{}-2\tau ^{AC\mu
}\nabla _{\mu }\psi _{A}^{B}{}.  \label{70}
\end{equation}%
Furthermore, calling for (\ref{45}) and (\ref{48}) along with the expansion (%
\ref{41}), after some computations, we get the contributions%
\begin{equation}
\check{D}^{A(B}\psi _{A}^{C)}{}=\Psi ^{ABC}{}_{M}\psi _{A}^{M}{}+\frac{2}{3}%
\varkappa \psi ^{BC}-\psi _{M}^{(B}{}\xi ^{C)M}  \label{71}
\end{equation}%
and%
\begin{equation}
\check{D}^{A[B}\psi _{A}^{C]}=M^{BC}\psi _{AM}\xi ^{AM}.  \label{72}
\end{equation}

We can see that the symmetry property of the wave functions entails
imparting symmetry in the indices $B$ and $C$ to the $\square $-block of (%
\ref{67}), which means that%
\begin{equation}
M^{A[B}\square \psi _{A}^{C]}{}=\frac{1}{2}M^{BC}M^{A}{}_{D}\square \psi
_{A}^{D}{}\equiv 0.  \label{73}
\end{equation}%
In both formalisms, Eq. (\ref{73}) thus implies that 
\begin{equation}
2\Delta ^{A[C}\psi _{A}^{B]}{}{}=M^{BC}(\frac{1}{2}m^{2}\nabla _{\mu }A^{\mu
}-\nabla _{\mu }s^{\mu }),  \label{74}
\end{equation}%
while the relations (\ref{70})\ and (\ref{72}) yield the expression%
\begin{equation}
\Delta ^{A[C}\psi _{A}^{B]}=M^{BC}(\tau _{M}^{A\mu }{}\nabla _{\mu }\psi
_{A}^{M}-\psi _{AM}\xi ^{AM}).  \label{75}
\end{equation}%
So, utilizing Eq. (\ref{29Lin}) and working out the $\tau \nabla \psi $-term
of (\ref{75}) to the extent that%
\begin{equation}
\tau _{M}^{A\mu }{}\nabla _{\mu }\psi _{A}^{M}=-[\frac{1}{2}\nabla _{\mu
}s^{\mu }+\nabla _{CB^{\prime }}(T^{AB^{\prime }}\psi _{A}^{C})+\psi
_{A}^{M}\nabla _{\mu }\tau _{M}^{A\mu }{}],  \label{76}
\end{equation}%
we arrive at the condition\footnote{%
When Eqs. (\ref{74})-(\ref{76}) are combined together, the terms that
involve $\nabla _{\mu }s^{\mu }$ explicitly get cancelled.}%
\begin{equation}
\frac{1}{4}m^{2}\nabla _{\mu }A^{\mu }+\nabla _{CB^{\prime }}(T^{AB^{\prime
}}\psi _{A}^{C})+\psi _{A}^{M}\nabla _{\mu }\tau _{M}^{A\mu }{}-\psi
_{A}^{M}\xi _{M}^{A}=0.  \label{77}
\end{equation}%
For $\phi _{A}^{B}$, we similarly obtain the massless condition%
\begin{equation}
\nabla _{CB^{\prime }}(T^{AB^{\prime }}\phi _{A}^{C})+\phi _{A}^{M}\nabla
_{\mu }\tau _{M}^{A\mu }{}-\phi _{A}^{M}\xi _{M}^{A}=0,  \label{78}
\end{equation}%
along with the complex conjugates of (\ref{77}) and (\ref{78}).

The property (\ref{73}) stipulates in either formalism that the only
contributions to the wave equation for $\psi _{A}^{B}$ are those produced by
the symmetric pieces in $B$\ and $C$ of the corresponding configuration (\ref%
{67}). Hence, carrying out a symmetrization over the indices $B$ and $C$ of (%
\ref{67}), likewise fitting together the pieces of Eqs. (\ref{69})-(\ref{71}%
) and rearranging indices adequately thereafter, we end up with the dark
energy equation%
\begin{equation}
(\square +\frac{4}{3}\varkappa +m^{2})\psi _{A}^{B}+2\Psi ^{LB}{}_{MA}\psi
_{L}^{M}=2\beta _{A}^{B},  \label{79}
\end{equation}%
with%
\begin{equation}
\beta ^{AB}=\nabla _{B^{\prime }}^{(A}s^{B)B^{\prime }}+\psi _{M}^{(A}\xi
^{B)M}+2(\nabla _{\mu }\psi _{M}^{(A})\tau ^{B)M\mu }+m^{2}\tau ^{AB\mu
}{}A_{\mu }.  \label{80}
\end{equation}%
We should emphasize that the statements (\ref{77})-(\ref{79}) are formally
the same in both formalisms, and additionally bear gauge invariance because
of the behaviour of $A_{\mu }$ as specified by Eq. (\ref{9}). Indeed, it is
the masslessness of the CMB fields that ensures the absence from (\ref{78})
of a term proportional to $\nabla _{\mu }\Phi ^{\mu }$.

It now becomes clear that the application to Eq. (\ref{79}) of the
correlations supplied by (\ref{62}) and (\ref{63}), allows us to attain
quite easily the $\gamma $-formalism version of the wave equations for $\psi
_{AB}{}$ and $\psi ^{AB}{}$. In effect, we have%
\begin{equation}
(\square -2i\alpha ^{\mu }\nabla _{\mu }+\overline{\Theta }{}+\frac{4}{3}%
\varkappa +m^{2})\psi _{AB}-2\Psi {}_{AB}{}^{LM}\psi _{LM}=2\beta _{AB}
\label{81}
\end{equation}%
and%
\begin{equation}
(\square +2i\alpha ^{\mu }\nabla _{\mu }+\Theta +\frac{4}{3}\varkappa
+m^{2})\psi ^{AB}-2\Psi ^{AB}{}_{LM}\psi ^{LM}=2\beta ^{AB},  \label{82}
\end{equation}%
which satisfy the rule (\ref{64}). For the $\varepsilon $-formalism, we
obtain%
\begin{equation}
(\square +\frac{4}{3}\varkappa +m^{2})\psi _{AB}-2\Psi {}_{AB}{}^{LM}\psi
_{LM}=2\beta _{AB}  \label{83}
\end{equation}%
and%
\begin{equation}
(\square +\frac{4}{3}\varkappa +m^{2})\psi ^{AB}-2\Psi ^{AB}{}_{LM}\psi
^{LM}=2\beta ^{AB}.  \label{84}
\end{equation}%
We notice that the $\varepsilon $-formalism lower-index version of $\beta
^{AB}$ is expressed simply as%
\begin{equation}
\beta _{AB}=\nabla _{B^{\prime }(A}s_{B)}^{B^{\prime }}-\psi _{(A}^{M}\xi
_{B)M}-2(\nabla _{\mu }\psi _{(A}^{M})\tau _{B)M}{}^{\mu }+m^{2}\tau
_{AB}{}^{\mu }{}A_{\mu }.  \label{84Lin}
\end{equation}

Due to the occurrence of the same formal geometric patterns on the
right-hand sides of the field equations of Section 3, we can promptly obtain
the CMB wave equations from the statements (\ref{79})-(\ref{84}) by first
setting $m=0$ and then replacing wave functions appropriately. In either
formalism, we thus have%
\begin{equation}
(\square +\frac{4}{3}\varkappa )\phi _{A}^{B}+2\Psi ^{LB}{}_{MA}\phi
_{L}^{M}=2\eta _{A}^{B},  \label{85}
\end{equation}%
with%
\begin{equation}
\eta ^{AB}=\nabla _{B^{\prime }}^{(A}\mathfrak{s}^{B)B^{\prime }}+\phi
_{M}^{(A}\xi ^{B)M}+2(\nabla _{\mu }\phi _{M}^{(A})\tau ^{B)M\mu }
\label{86}
\end{equation}%
and $\mathfrak{s}_{\mu }$ being given by (\ref{32Linn}). The $\gamma $%
-formalism equations for $(\phi _{AB},\phi ^{AB})$ accordingly appear as%
\begin{equation}
(\square -2i\alpha ^{\mu }\nabla _{\mu }+\overline{\Theta }{}+\frac{4}{3}%
\varkappa )\phi _{AB}-2\Psi {}_{AB}{}^{LM}\phi _{LM}=2\eta _{AB}  \label{87}
\end{equation}%
and%
\begin{equation}
(\square +2i\alpha ^{\mu }\nabla _{\mu }+\Theta +\frac{4}{3}\varkappa )\phi
^{AB}-2\Psi ^{AB}{}_{LM}\phi ^{LM}=2\eta ^{AB},  \label{88}
\end{equation}%
whereas the $\varepsilon $-formalism counterparts of Eqs. (\ref{87}) and (%
\ref{88}) are stated as%
\begin{equation}
(\square +\frac{4}{3}\varkappa )\phi _{AB}-2\Psi {}_{AB}{}^{LM}\phi
_{LM}=2\eta _{AB}  \label{89}
\end{equation}%
and%
\begin{equation}
(\square +\frac{4}{3}\varkappa )\phi ^{AB}-2\Psi ^{AB}{}_{LM}\phi
^{LM}=2\eta ^{AB}.  \label{90}
\end{equation}

\section{Concluding Remarks and Outlook}

The description we have just proposed here has been based upon the belief
that the spinor structures of generally relativistic spacetimes should
support locally a geometric description of the microwave and dark energy
backgrounds of the universe. Because of the fact that any torsional affine
potentials must always enter geometric prescriptions together with adequate
torsionless companions, we could definitely establish that any torsional
two-component spinor description of the dark energy background must be
accompanied by a description of the CMB. We saw that all wave functions
couple to the pieces of the spinor decomposition for the torsion tensor of $%
\mathfrak{M}$. They also interact with curvatures via couplings like, say,
the $\Psi \psi $ and $\Psi \phi $ ones carried by Eqs. (\ref{83}) and (\ref%
{89}). However, they do not interact with one another whence we can say that
one background propagates in $\mathfrak{M}$ as if the other were absent.
This result appears to be in full agreement with the suggestion made earlier
in Ref. [33] by which the CMB may propagate alone in spacetimes equipped
with torsionless affinities as Infeld-van der Waerden photons.

One of the striking aspects of the procedures implemented in Section 5, is
related to the gauge invariance of the condition (\ref{78}), which takes
place because the masslessness of the CMB fields annihilates either $\gamma
\varepsilon $-contribution that carries the non-invariant piece $\nabla
_{\mu }\Phi ^{\mu }$. It should be stressed that the occurrence of the
massive condition (\ref{77}) rests upon the torsionfulness intrinsically
borne by Eq. (\ref{75}). In the limiting case of the torsionless framework,
the derivative $\Delta ^{A[C}\phi _{A}^{B]}$ becomes an identically
vanishing contribution in both formalisms, and Eqs. (\ref{79})-(\ref{84})
all "evaporate" together with the source $\mathfrak{s}^{\mu }$ and the
curvature spinor $\xi _{AB}$. Under this circumstance, the world-spin scalar 
$\varkappa $ bears reality and satisfies the equality%
\begin{equation*}
4\varkappa =R,
\end{equation*}%
with $R$ being the Ricci scalar of $\nabla _{\mu }$. Hence, the
electromagnetic wave equations of Ref. [32] may be recovered, with the
physical significance described in Ref. [33] being of course effectively
ascribed to them.

We expect that the subsidiary conditions involved in the derivation of the
wave equations for the dark energy background could perhaps shed some light
on the physical meaning of the right-hand side of Einstein-Cartan's field
equations. We also believe that a distributional treatment of our wave
equations could be of considerable importance insofar as it may provide us
with local theoretical evaluations of the dynamical properties of dark
energy, including the feasibility of performing explicit calculations
towards making direct comparisons with data coming from the observed
anisotropy of the CMB. One of our hopes is that the role played by spacetime
torsion could be actually further strengthened. It is considerably
interesting to point out that the calculational techniques developed in
Section 4 can supply geometric tools for describing the propagation of
gravitons and Dirac particles in torsional cosmic environments, in
combination with the mechanisms that may avert gravitational singularities
as particularly exhibited in Ref. [43].

ACKNOWLEDGEMENT: I should like to thank the referees for making some
suggestions that have produced a significant improvement on the paper
presented here.


\begin{thebibliography}{99}
\bibitem{1} Adam G. Riess \textit{et al}., Astro. Jour., 116, 1009 (1998).

\bibitem{2} Saul Perlmuter \textit{et al}., Astro. Jour., 517, 565 (1999).

\bibitem{3} Philip J. E. Peebles and Bharat Ratra, Rev. Mod. Phys. 75, 559
(2003).

\bibitem{4} Edmund J. Copeland \textit{et al}., Int. J. Mod. Phys. D15, 1753
(2006).

\bibitem{} Albert V. Minkevich, Phys. Lett. B, Vol. 678, Issue 5, 423 (2009).

\bibitem{} Thomas Buchert, Gen. Rel. Grav., 40, 467 (2008).

\bibitem{} Christian Beck and Michael C. Mackey, Int. J. Mod. Phys. D 17, 71
(2008).

\bibitem{} Arthur D. Chernin \textit{et al}., Astrophys. 50, 405 (2007).

\bibitem{9} Aragam R. Prasanna and Subhendra Mohanty, Gen. Rel. Grav., 41,
1905 (2009).

\bibitem{} Charles L. Bennett \textit{et al}., Astrophys. J. Supp. Series
148, 1 (2003).

\bibitem{} David N. Spergel \textit{et al}., Astrophys. J. Supp. Series 148,
148 (2003).

\bibitem{} Thanu Padmanabhan, Physics Reports 380, 235 (2003).

\bibitem{13} Christian G. B\"{o}ehmer and Tiberiu Harko, Eur. Phys. J. C50,
423 (2007).

\bibitem{} Andrzej Trautman, Encyclopedia of Mathematical Physics, Vol. 2,
189 (Ed. by J. P. Fran\c{c}oise, G. L. Naber and S. T. Tsou, Oxford,
Elsevier 2006).

\bibitem{} Friedrich W. Hehl \textit{et al}., Rev. Mod. Phys., Vol. 48, No.
3, 393 (1976).

\bibitem{16} Roger Penrose and Wolfgang Rindler, Spinors and Space-Time Vol.
1, (Cambridge University Press, Cambridge 1984).

\bibitem{} Tom W. Kibble, Jour. Math. Phys., 2, 212 (1961).

\bibitem{18} Dennis W. Sciama, \textit{On the Analogy Between Charge and
Spin in General Relativity} In: Recent Developments in General Relativity
(Pergamon and PWN, Oxford 1962).

\bibitem{} Nikodem J. Poplawski, Phys. Lett. B, Vol. 694, No. 3, 181 (2010).

\bibitem{} Christian G. B\"{o}ehmer, Acta Phys. Polon. B , 36, 2841 (2005).

\bibitem{21} Ilya L. Shapiro, Physics Reports 357, 113 (2002).

\bibitem{} Venzo De Sabbata, Astrophysics and Space Science Vol. 158, 347
(Kluwer Academic Publishers, Belgium 1989).

\bibitem{} Salvatore Capozziello \textit{et al.}, Eur. Phys. Jour. C, 72,
1908 (2012).

\bibitem{} Thomas P. Sotiriou and Stefano Liberat, Ann. Phys., 322, 935
(2007).

\bibitem{25} Thomas P. Sotiriou and Valerio Faraoni, Rev. Mod. Phys., 82,
451 (2010).

\bibitem{26} Leopold Infeld and Bartel L. Van der Waerden, Sitzber. Akad.
Wiss., Physik-math. Kl. 9, 380 (1933).

\bibitem{} Hermann Weyl, Z. Physik 56, 330 (1929).

\bibitem{} Jorge G. Cardoso, Czech Journal of Physics 4, 401 (2005).

\bibitem{29} Jorge G. Cardoso, Adv. Appl. Clifford Algebras 22, 985 (2012).

\bibitem{} Roger Penrose, Ann. Phys., 10, 171 (1960).

\bibitem{} Louis Witten, Phys. Rev., 1, 357 (1959).

\bibitem{32} Jorge G. Cardoso, Acta Phys. Polon.,Vol. 38, 8, 2525 (2007).

\bibitem{33} Jorge G. Cardoso, \textit{Wave Equations for Invariant
Infeld-van der Waerden Wave Functions for Photons and Their Physical
Significance}, In: Photonic Crystals, Optical Properties, Fabrication ...,
Editor: William L. Dahl 2010 Nova Science Publishers Inc., New York (ISBN
978-1-61122-413-9).

\bibitem{34} Herbert Jehle, Phys. Rev. 75, 1609 (1949).

\bibitem{} William L. Bade and Herbert Jehle, Rev. Mod. Phys., Vol. 25, 3,
714 (1953).

\bibitem{} Peter G. Bergmann, Phys. Rev., Vol. 107, \textit{2}, 624 (1957).

\bibitem{} Ezra T. Newman and Roger Penrose, Jour. Math. Phys. \textit{3},
566 (1962).

\bibitem{38} Roger Penrose and Wolfgang Rindler, Spinors and Space-Time Vol.
2, (Cambridge University Press, Cambridge 1986).

\bibitem{39} Jorge G. Cardoso, European Physical Journal Plus Vol. 130, 
\textit{1}, 10 (2015).

\bibitem{40} Jerzy Plebanski, Acta Phys. Polon. 27, 361 (1965).

\bibitem{41} Roger Penrose, Foundations of Physics, Vol. 13, No. 3, 325
(1983).

\bibitem{42} Andrzej Trautman, Symposia Mathematica, 12, 139 (1973).

\bibitem{43} Andrzej Trautman, Nature (Phys. Sci.), 242, 7 (1973).

\bibitem{44} Andrzej Trautman, Annals N. Y. Acad. Sci., 262, 241 (1975).

\bibitem{45} Robert Geroch, Jour. Math. Phys. 9 (1968) 1739.

\bibitem{46} Robert Geroch, Jour. Math. Phys. 11 (1970) 343.

\bibitem{47} Mark Burgess, \textit{Classical Covariant Fields, }Cambridge
Monographs on Mathematical Physics (Cambridge University Press, Cambridge
2005).
\end{thebibliography}
\end{document}